# Introduction to ethics in the age of digital communication

Authors: Rebekah Rousi and Ville Vakkuri

## Introduction

The field of ethics, or moral philosophy (scholarship of moral principles and codes), has been ongoing throughout human history. As with any area related to the field of philosophy, a precise definition of 'ethics' does not exist (Hitt, 1990). Rather, general understanding of what ethics are can be described simplistically as the study of what is 'good and bad' or 'good and evil' (Singer, 2023). More relevant for today's societal discourses, are behavioural understandings of ethics, and ethical practice (Barraquier, 2011; Watson, 2007). When returning to a basic behavioral level understanding of ethics we may refer to the Golden Rule, "[i]n everything, do to others what you would have them do to you..." (New International Version Bible, 2011, Matthew 7:12). Thus, a rule-of-thumb in terms of ethical conduct can be considered as guiding one's behaviour in terms of how one would like to be treated (Saariluoma & Rousi, 2020). On a daily professional and academic basis, considering ethical issues and even interpreting, as well as complying to ethical standards may be more challenging than it appears. From this perspective, it is integral for individuals operating in the fields of communications to grasp what ethics are, how they relate and apply to specific domains, where basic principles or similarities lie from context to context and where there may be differences.

In communication, professionals, researchers and students are obligated to comply with numerous ethical standards or rules within their work. In addition to the external pressures of complying with ethical standards or rules, practitioners should possess an intrinsic sense of ethical responsibility that is instilled and maintained through reflective practice (Hargreaves & Page, 2013). As communication experts, individuals should engage in activities in a way that holds the most benefits and least harm for all involved.

From a temporal-cultural perspective for instance, ethics and understandings of what is 'right or wrong' continuously evolve (Hand, 2018; Singer, 2011b). There are numerous factors influencing interpretations of right and wrong, good or bad, in terms of human conduct that either gradually or rapidly emerge in cultures over time. Characteristics of multiculturalism, interculturalism or globalisation automatically imply a plurality of cultures (Miike, 2019). This entails that ethical framing or the definition of moral codes in turn becomes multi-faceted and potentially entangled (Wakkary, 2021; Zhong et al., 2022). By entangled we refer to the intertwined nature of multiple cultural, social and indeed ethical perspectives that manifest and are subject to compliance, negotiation or compromise depending on context or purpose (Tavory, 2022). As stated by Lenard (2020), "it is important to understand the role of culture in not only what is valued but in how practices are to be morally evaluated." Moral evaluation and indeed ethical engagement involve critical thinking (Paul & Binker, 1990). Critical thinking entails active analysis of phenomena from multiple perspectives in order to derive a conclusion on the phenomena's ethical weighting in light of values and moral codes (Ennis, 1991).

In the era of global connectivity individuals are constantly negotiating several sets of ethical understandings via interactions with cross-institutional and international contacts and colleagues on a daily basis (Casmir, 2013). Additionally, entangled with globalisation are economic, technological and political factors. These escalate ethical complexity and in themselves affect societal definitions, illustrations and implementations of ethical matters and moral codes of conduct. There are many examples that can be observed in terms of how these codes of conduct have changed, one such being the acceptance of curse (swear) words in Western mass media over recent decades. Until the 1960s the over 550 year old 'F' (*fuck*) word was largely invisible in mass media. With the exception of statement makers such as the Guardian Editor Alistair Hetherington, who allowed the word to be published in relation to an article on Lady Chatterley in 1963 (Collins, 2015), the word was more released in institutional communication. Yet, over the past decades, changes can be observed in public television whereby this word, and others that are similar have been bleeped and censored.

Similarly, socio-technological developments such as the rise of social media and dating applications (apps) have contributed to changes in the ways people not only respond to language, but also view changes in relationship structures and what is considered appropriate behaviour. An app like Tinder for instance, whilst framed as a dating app is noted for its role in facilitating ephemeral intimate encounters (Rousi, 2021). Interestingly, where several decades ago (i.e., 1990s), Internet facilitated romantic connections were viewed with high levels of scrutiny (Katz & Aspden, 1997), these days both Internet-mediated romantic connections and short-lived affairs are normalised in Western societies (Castro & Barrada, 2020). This has been the result of both social movements as well as techno-economic - *if sex sells, then why not? Money talks.* These issues are often categorised as relating to moral standards and advances in social norms (what is acceptable and what is not). However, at the heart of decisions regarding whether to push moral boundaries, and how we respond to the breaking of boundaries, additionally relates to ethical questions particularly since human wellbeing (emotional, social, spiritual, economic etc.) may be at risk.

Social and linguistic matters are not the only burgeoning ethical areas in the field of communication. Social-cultural power relations, privilege and bias (of voice, representation, and producer-consumer balance, ownership etc.), censoring and freedom of speech are all issues that are not becoming easier as our technology advances (Kaplan & Haelein, 2020). This adds dimensions to communication ethics specific to technologies such as artificial intelligence (AI) for instance. These dimensions will be discussed towards the end of this article. Given the influence of communication, and its intricate connection to emerging and future technology, it is important for future practitioners and researchers to at least hold a basic understanding of ethics, how they relate to both communication (i.e., communication ethics, business ethics, consumer ethics) and technology (i.e., techno-ethics, AI ethics, information ethics etc.). Moreover, from the perspective of the media, most broadcasting companies hold their own code-of-ethics by which employees should adhere (Wulfemeyer, 1990). These branches of ethics belong to the field of applied ethics. Applied ethics refers to ethical considerations of real-world problems (Beauchamp, 2005). This renders applied ethics particularly important for communication specialists.

The article begins with a brief summary and history of applied ethics as a practical sub-field of ethics. It then moves on towards observing ethics in contemporary professional practice from both a practical perspective, as well as in relation to some of the ways they have been studied. The article then discusses the ways in which the nature of ethics in the field of communication have been changing, and the impact of emerging technology on these changes.

## Categorical description of applied ethics

Ethics can be described as a philosophical field of study (see e.g. Dittmer, 2013.). In particular, ethics is the study of morality. As a field of research, ethics can be organized into three subfields: metaethics, normative ethics, and applied ethics (see Figure 1). Metaethics focuses on the nature of moral truths and the character of morality itself. Normative ethics focuses on moral theories and principles. Traditionally normative ethics is divided into subfields deontological ethics, consequentialism, and virtue ethics. Deontological ethics focuses on intention, and fundamental rights or moral duties. Consequentialism centers on outcomes in terms of overall goodness, and what the achievement of this goodness would mean from the perspective of acceptable moral behaviour. Virtue ethics concentrates on virtuous (high integrity and demonstration of highly moral behaviour) character and what needs to be done in order to achieve or create virtuous character (see e.g., Audi, 2012; Borden, 2013).

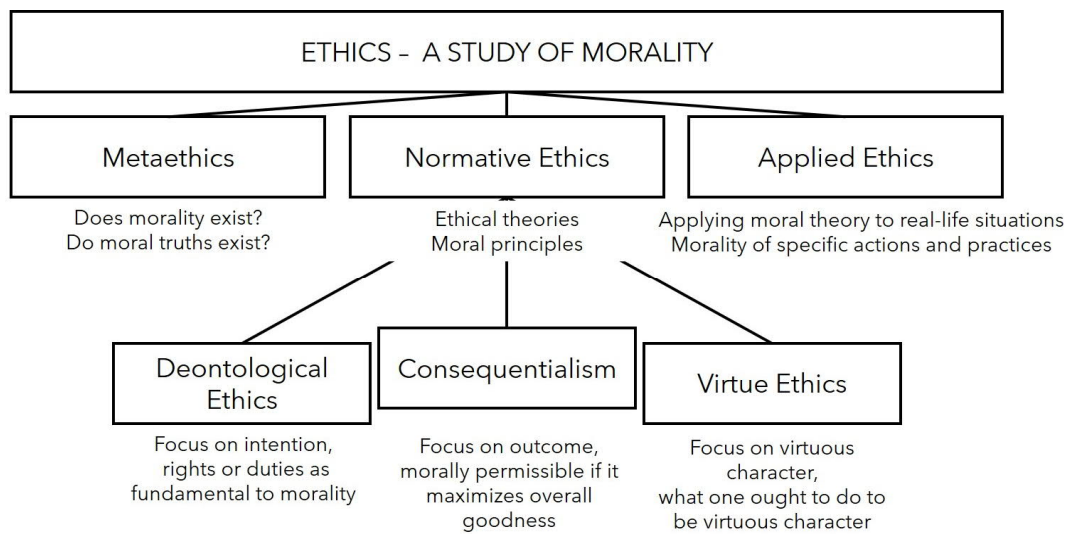

**Figure 1**. Subfields of ethics

The role of applied ethics is to examine unclear and murky (or wicked) real-life situations to understand what would be permissible-impermissible, or right-wrong actions. Due to its nature, questions of applied ethics are more specific and situational than those of metaethics or normative ethics (Dittmer, 2013). For example, specific questions like: What are the social responsibilities of companies?; Why should companies care about the environment?; Are some tangible considerations included in the applied ethics domain. Applied ethics often enters highly complicated fields of problems, meaning that seldom one is left with a unified or unproblematic solution. Subfields of applied ethics include business (see e.g., DesJardins, 2009), bio (see e.g., Engelhardt, 1996), professional (see e.g., Koehn, 2006), social (see e.g., Downie, 2020), and technology (see e.g., Tavani, 2003), information (see e.g., Floridi, 1999), and AI ethics (see e.g., Heilinger 2022).

The subfields of applied ethics are highly relevant for the area of communication in research and practice for many reasons. Reasons pertain both primarily to communication itself and issues related to communication (including for instance, communication technology and mediation) as well as to the issues that communication deals with (mediating the message). This impacts *what* is communicated, *how* it is communicated, *who* is communicating, and *who* are the receivers - the intended receivers of the messages and the unintended receivers.

Another way to think of applied ethics is that of 'ethics as practice' (see e.g., Barraquier, 2011; Watson, 2007). Ethics as practice is as an academic endeavor to study and understand situational real-life problems as well as an effort to guide practice to help individuals or communities achieve justified moral stances regarding specific issues. For example, professional codes-of-ethics are efforts to implement ethical theory in action. If the focus on applied ethics is situations and scenarios from real-life, then it is relevant to ask which of these situations are morally relevant. Central questions driving ethical inquiry relate to considerations for what is ethical, unethical and even non-ethical. Thus, considerable effort is also placed on understanding what actually qualifies as an ethical issue, and what does not, especially in light of the nature of ethical issues. Other questions include: whether or not everything can be considered to entail ethical issues; what actions or behaviour can be considered as ethical; and whether or not all ethical matters and solutions are always debatable at some level. One approach to help understand this is the categorisation actions to non-ethical, ethical and unethical (see Figure 2).

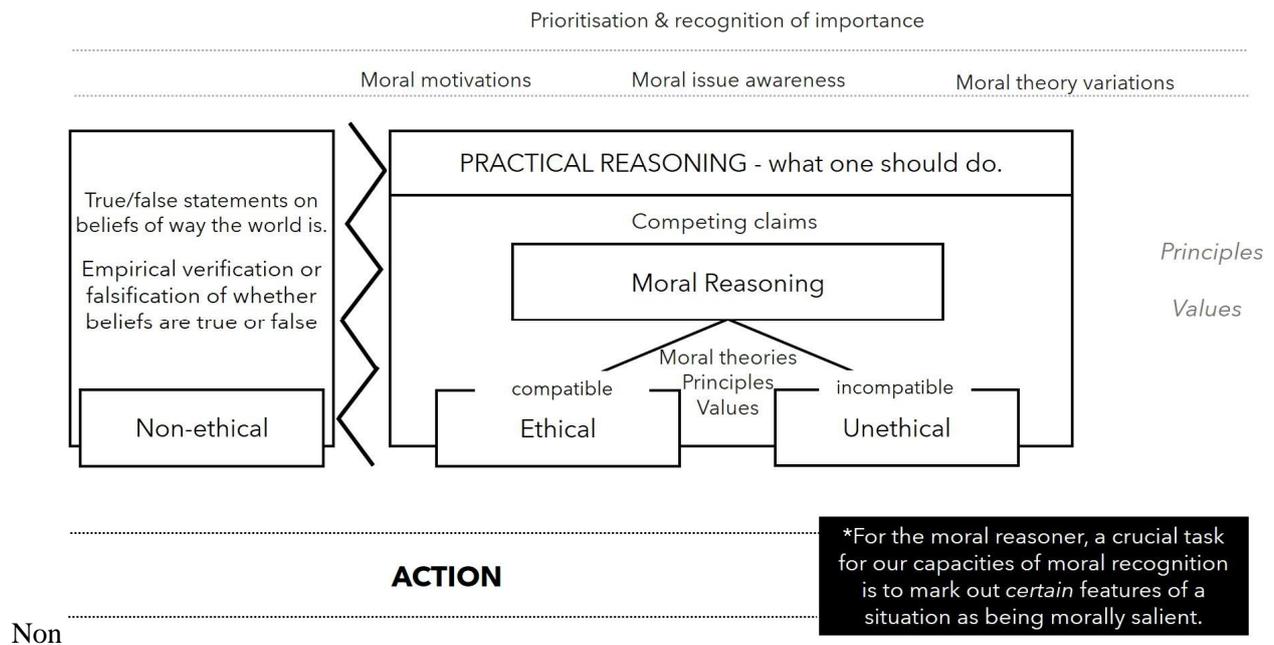

Non

**Figure 2.** Non-ethical, ethical and unethical matters

The non-ethical, requires that questions are asked regarding true or false statements. These true or false statements inquire about world view, and how people believe that the world operates or exists (see e.g., Dreier, 2004; Wallace, 2020). For instance, so-called truths that may be claimed are that this article is written in English, or that 2+1=3, or that water freezes under zero-degree celsius etc. Within this domain, efforts are made to either support these claims or reject them. In contrast to true or false statements, practical reasoning deals with more open questions. These questions may pertain to what one ought to do in given situations. For example, how can one get from point A to point B, or what one would need to do in order to buy soda (Singer, 2011a). Moral reasoning is deliberated according to competing claims. Often these claims are connected to discourse that is contextually-situated. The ethical (compatible) and unethical (incompatible) are defined by moral theories, principles and values that are utilised to justify particular stances. (Wallace, 2020) One ready example can be seen in Facebook's recent choice to allow hate speech and imagery against the Russian military in relation to the Ukraine invasion - i.e., Russia has invaded the Ukraine and is causing pointless damage, hurt and deaths, *thus*, we will allow the expression of negative sentiments on our social media platform[1].

## Ethics and technology in the field of communication

While surprisingly seldom acknowledged, and often undervalued among other professions, the field of communication may be deemed one of the most crucial regarding the shaping of worldviews (Pentland, 2010). The ways that issues are discussed, written about and framed, influence how they are understood (Schroeder & Borgerson, 2005). Rhetoric (e.g., Herrick, 2020), discourse (e.g., Fairclough, 1992), framing (e.g., Goffman, 1974.), intertextuality (e.g., Orr, 2010), and intermediality (e.g., Allen, 2021) all serve to create an image of phenomena that portrays other underlying messages. As one African proverb states:

> *"Until the lions have their historians, tales of hunting will always glorify the hunter."*
> (Oxfam Poster, London, 1989, cited in Fore, n.d.)

Communication is tightly intertwined with disciplines such as linguistics, literature, media and journalism - all of which engage in communication practice in various ways (Beacco et al., 2002; Richards & Schmidt, 2014). In recent decades, communication has steadily integrated diverse knowledge from other fields such as cultural studies and philosophy, to cognitive science, psychology and even computer science (Walton, 2000). Communication relies on and reinforces culture, which is manifested through technology as a means of practice (Androutsopoulos, 2011). Both culture and technology express and affect the human mind (Boguslawska-Tafelska, 2013; Kuhl et al., 2001).

---

[1] For more information on this issue see e.g., Veganttil & Culliford (2022).

As technology grows in complexity, so do the ethical challenges. Technology-related ethical issues are vast. Here, we will focus on the examples of misinformation, disinformation and deep fakes that operate across levels of information, from what seems like public news media to apparent personal communication between associates (Shu et al., 2020). Misinformation refers to information that is not correct, while disinformation is the intentional use or construction of incorrect information motivated by various agenda (Scheufele & Krause, 2019), such as scare campaigns, or malicious attacks on reputation. Deep fake technology is driven by AI that morphs videos enabling modifications of video content and particularly people within that content to say and do actions in a hyperealistic way (Westerlund, 2019). Social media is flooded with deepfakes, making it difficult to distinguish authentic content and identities from fakes[2]. Deep fakes affect information pathways such as news media, and disrupt the credibility of other communication domains, e.g., advertising (Campbell et al., 2022). Deep fakes reduce certainty in organisations, eroding trust at even societal levels (Vaccari & Chadwick, 2020).

There is another dimension to the intricate relationship between communication and technology, and that is the ways in which communication manifests, frames, and promotes technology - its need, its nature (ethical - good or bad), and its timeliness (see Figure 3).

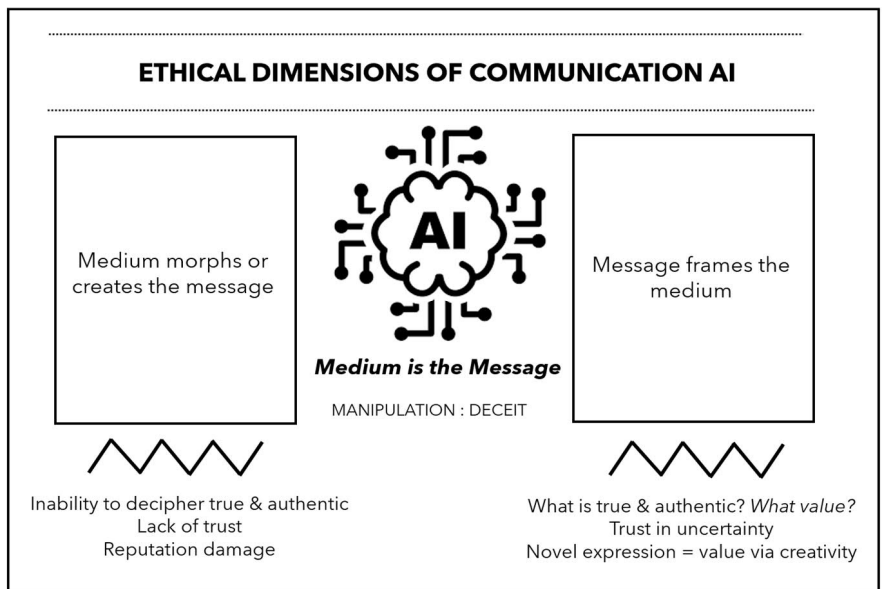

**Figure 3.** Ethical dimensions of communication AI - morphed by technology and morphing technology

Figure 3 illustrates the ethical dimensions of AI in communication through the understanding that technology both shapes communication and is shaped by communication. The diagram directly implies the dynamics of AI in communication, particularly in light of mis- and dis-information and deep fakes. Moreover, the confounding effects of deceit and manipulation within encountered AI-driven communication may lead to the questioning and appropriateness of the use of AI as a medium for communication as a whole. Potentially, when understanding the right-hand side of the figure, doubt may be shed on the very existence of truth or authenticity via the understanding that all communication is mediated by human beings. The value of perceived truth could also be interrogated in an information society whereby survival of the loudest and proudest with the most convincing reasoning is paramount. One ethical twist to the case of deep fakes are their creative value. While unethical is posed as genuine, the cultural products themselves boast creativity and technological ingenuity (non-ethical).

---

[2] For more information on the logic of deep fake technology see, Mike Boyd's "I learned to make deep fakes… and the results are terrifying" (2023; https://www.youtube.com/watch?v=Zmutd9618Kk)

# Conclusion - ethics in practice for communication in the era of AI

*"The threat of Stalin or Hitler was external. The electric technology is within the gates, and we are numb, deaf, blind, and mute about its encounter with the Gutenberg technology, on and through which the American way of life was formed."*

(McLuhan & Fiore, 1967/2006, p. 114)

The ultimate ethical concern of communication in itself, is its capacity to influence and describe the state of the human mind (Meltzoff, 1999). McLuhan and Fiore (1967) could see (to some extent) the difference between the war from outside (physical violence) to the war from within. Our societies have historically been, and are now more than ever, controlled by the ways in which information structures these societies. We have rapidly moved from the printing press producing physicalised (analogue) communication artefacts (books, newspapers etc.) - yet still, equally as capable of quantifying falsehoods and fantasy - to an era in which physicality (atoms) transfer into media (bits). It is not simply the words that penetrate our thoughts, but an ensemble of multimedia that takes shape both through human agency (the communicator - communication professional and those behind them) and through the technology that curates it. The example of this article has been of AI.

At the heart of ethics, ethical choices, and moral principles, is the will and intention to do what is right (Nietzsche, 2004). There is the need to understand what is right or wrong, and even more, understand the complexity of right (good) and wrong's (bad/evil) definition (Narvaez, 2010). The greatest challenge of this technological era is the information intensive nature of the technology itself. Problems stemming from these include: 1) power of those who hold and control information (via technology - ownership); 2) cultural-linguistic privileges in information types, how it is available, who has access, and how *via ethics* certain populations (cultures and sub-cultures) and their behaviours are labelled as good or bad; 3) the growing gap between those who are tech-savvy, understanding the inner workings of the technology, and those who do not; and 4) the growing uncertainty, not only of credibility in the stories and representations encountered in the media (online and otherwise), but in one's own ability to decipher what is true or not, leading to doubts in one's sanity and cognitive abilities such as memory. Through applying ethical principles both from the field of communication (see e.g., Fox et al., 2021) and technology-related areas (see e.g., Vakkuri et al., 2021; Vargas-Elizondo, 2020), communication professionals and researchers are able to mitigate the ethical grey areas promoted by the technology and ill-willed users, through their own practice and integrity.